This is the accepted version of the following document:

Leonardo Leyva, Daniel Castanheira, Adão Silva, Atílio Gameiro, and Lajos Hanzo, "Cooperative multi-terminal radar and communication: A new paradigm for 6G mobile networks," accepted for publication in *Vehicular Technology Magazine (VTM)*.

# Cooperative multi-terminal radar and communication: A new paradigm for 6G mobile networks

Leonardo Leyva, Daniel Castanheira, Adão Silva, Atílio Gameiro, and Lajos Hanzo

*The impending spectrum congestion imposed by the emergence of new bandwidth-thirsty applications may be mitigated by the integration of radar and classic communications functionalities in a common system. Furthermore, the merger of a sensing component into wireless communication networks has raised interest in recent years and it may become a compelling design objective for 6G.*

*This article presents the evolution of the hitherto separate radar and communication systems towards their amalgam known as a joint radar and communication (RADCOM) system. Explicitly, we propose to integrate a radio sensing component into 6G. We consider an ultra-dense network (UDN) scenario relying on an active multistatic radar configuration and on cooperation between the access points across the entire coverage area. The technological trends required to reach a feasible integration, the applications anticipated and the open research challenges are identified, with an emphasis on high-accuracy network synchronization. The successful integration of these technologies would facilitate centimeter-level resolution, hence supporting compelling high-resolution applications for next-generation networks, such as robotic cars and industrial assembly lines.*

## Introduction

There has been significant interest in the combination of radar and communication functionalities within a common platform [1]-[4], and this need will increase with 6G, which aims for supporting the convergence of the digital, physical, and personal domains. This requires expanding the functionalities of 5G, to include the integration of radio sensing and communications in support of both hardware-and spectrum-sharing. The objectives of this article are to provide an overview of the current trends, the applications, and a path for their convergence. Finally, we propose a framework for the integration of a sensing component into an ultra-dense network (UDN).

## Radar and Communication Paths over Time

Radio detection, ranging and wireless communication are the most common radio functionalities used by both civilian and military applications. Traditionally, these systems have been designed and developed in isolation from each other [1]. Reflectometry-based sensing systems send a known waveform and assess the response of the environment. In radar systems, the transmitted waveform is reflected by targets, which facilitates the estimation of parameters, like the range, velocity, and angle [2]. By contrast, in communication systems, there is a need to estimate the particular parameters of the channel using sophisticated signal processing techniques. This is typically the channel's impulse response (CIR), which directly affects the transmitted information-carrying signal.

## Why is it Time to Merge Radar and Communication?

The independent design of these systems wastes valuable spectral resources. Therefore, the spectrum shortages combined with the emergence of novel applications requiring both functionalities, has inspired the research of joint radar and communication systems

The mobile data traffic growth may only be accommodated by higher-throughput systems using wider bandwidth. This trend has led to the shortage of spectrum, which is more evident in the sub-6 GHz bands. Novel areas considered for 6G, such as intelligent transportation systems (ITSs) and reflectometry-based Internet of Things (IoT) relying on sensors for the support of smart cities, could benefit from the application of a joint sensing-communication paradigm [3]. For instance, ITSs require smart cars to be capable of individually sensing the driving conditions and cooperatively sharing this information through a wireless communication network with other vehicles. Since sensing and communication are intrinsic functions of IoT devices, the convergence of both features in sensors based on reflectometry principles would be a critical step in providing dual functionality, promoting convergence and reducing the cost of production by reusing technology [3].

## The Road to the Merger of Radar and Communication

The protection against intersystem interference in radio has usually been guaranteed by the fulfilment of the regulatory directives defined by national and international bodies. The need for increasingly more spectrum has led to novel alternatives in spectrum regulations as well as to research on interference-management techniques and concepts, allowing the radio resources to be employed more efficiently [2]. Therefore, RADCOM systems have been the subject of significant research in recent years [3][4]. The coexistence of radar and communication systems, their degree of integration, and how interference management is implemented are summarized at a glance in Table 1.

The traditional coexistence considers the scenarios where radar and communications operate as totally independent systems. Here, regulators mainly allocate distinct bands for ensuring that there is no interference. However, if the same spectrum is used by different services, the operation must be meet protection (e.g., coordination zones may be enforced). In this scenario, the transceivers fulfil their mission in isolation from each other [2].

In the cognitive coexistence approach, radar and communications are still distinct systems serving different purposes, but in contrast to the previous category, they may temporally or permanently share the same frequency band more efficiently. Spectrum sharing is facilitated by cognitive radio techniques, where one of the services may use the band allocated to the other, provided that it does not inflict interference. This approach requires the sensing of spectral activity in order to check which bands are free (temporarily or geographically) and then to allow transmission over these bands [5]. There may be different ways of implementing the cognitive radio concept. The occupancy status of the bands may be in a regularly updated database, or the systems may rely on spectral sensing mechanisms to check the status [4]. Other coexistence approaches rely on projection-based methods [6], for which the transmit precoder of the secondary system is designed to reduce or avoid interference onto the legacy system. These methods usually require the estimation of the interference channel, which lead to increased complexity and signalling overhead. A more detailed and concise overview of these methods can be found in reference [6]. These more efficient approaches have been recently implemented for mitigating the spectral congestion problem.

The co-design approach breaks away from how spectral resources have traditionally been accessed, since the communication and radar systems are no longer treated as independent entities or as sources of interference [1][2]. Instead, their common elements, such as antenna arrays, radio frequency (RF) front-end circuits and digital signal processing (DSP) blocks, are shared [4], leading to improvements in their cost, form-factor, spectral efficiency and power consumption [3]. The convergence of radar and communication functionalities through co-design will be of interest to radar systems, which will be able to provide extra functionality based on transmitting information. Furthermore, it may also be motivated by the need to integrate the ability to sense the environment in 6G wireless networks.

*Organization*

The remainder of this article is organized as follows. The following section presents the different RADCOM techniques and classifies them according to their main functionalities. We continue by introducing a RADCOM framework relying on a sensing component in an UDN. Next, their enabling technologies, applications and implementational challenges are discussed. Finally, we provide our concluding remarks.

**RADCOM Systems**

As previously stated, the co-design may follow two different paths: relying either on a communication functionality or a radar component as its basis. This section is structured following this classification. First, we briefly review the integration of the communication functionality into radar systems. Then, we describe the main technologies already used in communications to jointly perform sensing and communication.

*Embedding Data into Radar Waveforms*

Regarding radar, information embedding is the main path pursued for convergence, which exploits the available hardware for communication [4][7], including the high-gain antennas and the radio front-end. Furthermore, an increased bandwidth and transmission power become available.

The authors of [7] discuss several information embedding techniques, mapping the information to be conveyed to either the phase, the amplitude, the frequency, or their combinations. Furthermore, beampattern modulation techniques [1][6], which modify the side-lobe levels (SLLs) in an opportunistic manner, provide another option to embed data into radar waveforms. Since these waveforms were optimized for the radar functionality, they achieve reliable sensing. However, they lead to a low data rate for communications, which is a severe limitation. For example, the resolution achieved by a pulsed continuous wave (CW) signal is 3.75 cm for a 4 GHz bandwidth. This resolution is suitable for a large number of applications, but a dedicated communication system having the same bandwidth achieves significantly higher data rates [8]. A compelling application would be autonomous driving, but the resultant data rate may become inadequate for reliable coordination among vehicles, especially in the face of the increased interference.

| APPROACH | SYSTEMS INTEGRATION LEVEL | INTERFERENCE MANAGEMENT | OBSERVATIONS |
|---|---|---|---|
| TRADITIONAL COEXISTENCE | No integration, i.e., radar and communication operate independently | Granted by regulatory bodies by the assignation of exclusive bands or protection methods (e.g., coordination zone) | No spectrum overlap with some exceptions. In some specific cases same frequency bands may be assigned to both systems, but not shared. |
| COGNITIVE COEXISTENCE | Limited integration, one or both systems may be aware of other transmission | Granted by cognitive techniques (e.g., opportunistic spectrum access and projection-based methods) and regulatory bodies | Increased spectral efficiency than the traditional coexistence at cost of complexity |
| CO-DESIGN | Total integration, designed to jointly perform radio-sensing and communication functions | Functional coexistence without interference, waveform and hardware are shared | Need to jointly fulfill radar and communication performance |

**Table 1** *Coexistence approaches for radar and communication systems.*

## Sensing Function Through Typical Communication Waveforms

To simultaneously achieve a high sensing performance and high data rates, novel waveforms are needed. The ones that are emerging reuse some concepts of communication systems and fall into two main categories: single carrier code-based and multicarrier-based waveforms [6]-[9].

In the single carrier code-based waveforms category, most of the implementations use a direct-sequence spread spectrum (DSSS), where the data sequence (in communications) is spread by a unique, user-specific pseudorandom sequence that allows for unambiguously distinguishing the users. This type of waveform has been used in wireless communication systems for a long time (e.g., IEEE 802.11g).

Several of its beneficial properties make single carrier code-based waveforms also suitable for radar applications [8]. Specifically, the design of sequences having good autocorrelation properties facilitates high-accuracy range estimation, while good cross-correlation allows for multiple radars to be active in the same band without excessive levels of interference [8]. A code-based multiple-access method, known as non-orthogonal multiple access (NOMA) [10], is being considered for next generation networks, which is equally suitable for both functionalities.

Despite these advantages, single carrier waveforms tend to be prone to Doppler effects [8]. Furthermore, spreading sequences can be designed for good cross-correlation properties, but their performance is often eroded in asynchronous systems and it is always limited by the Welch bound [8].

Multicarrier waveforms constitute a flexible candidate for joint radar and communication systems. These waveforms are the basis of several communication systems - e.g., IEEE 802.11 a/g/n, Long Term Evolution (LTE), and 5G New Radio (NR) - and have also become popular for radars. In [7]-[9], the authors showed that orthogonal frequency division multiplexing (OFDM) waveforms are adequate for radar applications, since they lead to improvements in parameter estimation (e.g., Doppler effect and range), target detection, spectral efficiency and radar resolution [9]. The use of multiple subcarriers improves the resource allocation flexibility compared to single carrier waveforms. Additionally, using multiple subcarriers and sophisticated DSP techniques [8] in the frequency domain allows the relaxation of the autocorrelation requirements when compared to single carrier waveforms [7].

But naturally, multicarrier waveforms also present a few drawbacks, given their high peak-to-average power ratio (PAPR), sampling rate and implementation complexity [8].

## The Cooperative Multistatic Paradigm: Sensing Capabilities for Next-Generation Networks

Future networks will require a sensing component for supporting the IoT, ITSs and smart cities, but the electromagnetic waves used for communication could also be used for radar-type sensing [11][12].

For example, OFDM has been shown to be suitable for joint radar-sensing and communications purposes. Moreover, wireless communication systems are migrating to the mmWave band, using frequencies like those employed in radar. The generic concept of *Perceptive Mobile Networks* was proposed in [11], which includes operation based on monostatic and bistatic radar philosophies. The issue with the integration of traditional monostatic radar solutions into a cellular network is the need for full duplex transceivers, but this technology is far from maturity. The full duplex requirement led to solutions based on the bistatic radar concept advocated in [12] for automotive applications.

In this article we propose a practical solution for the radio sensing component to be integrated into the 6G networks. More specifically, we propose the concept of a cooperative multistatic radar intrinsically embedded into the wireless cellular network architecture. In cooperative multistatic radar only the fixed elements of the network are involved in the radio sensing process, while in the most generic cooperative multi-terminal concept, it may involve mobile network elements. The proposed cooperation allows a reduction of the bandwidth required, since through the combination of multiple links we get an effect similar to the capacity enhancement of 'massive' multiple-input multiple-output (MIMO) systems. However, having high-accuracy synchronization becomes absolutely vital [13], since if the access points are not both tightly time and phase-synchronized, the spatial resolution attained degrades. For a perfectly synchronized network, the spatial resolution accuracy will be inversely proportional to the carrier frequency. By contrast, for an unsynchronized one, it is inversely proportional to the typically much lower signal bandwidth [13]. In the latter case, the cooperation between the network elements fails to improve the resolution, since it would be identical to that of a monostatic radar. However, time synchronization alone is insufficient for improving the resolution, even when the cooperation of network nodes is enforced. Time synchronization together with cooperation leads to a more accurate estimation of the object's position. This is because more measurements become available and the spatial variability of the radar cross section (RCS) improves the radar detection performance. Still, the resultant resolution remains lower than in the presence of phase-and time-synchronization.

We continue by describing the proposed system architecture, followed by briefly introducing the main technologies necessary for constructing a functional RADCOM network. Finally, the applications foreseen and open research challenges are presented.

*System Architecture: General Overview*

Figure 1 shows the RADCOM architecture proposed for the integration of a sensing component into next-generation networks. The scenario considered is an UDN, where the high number of base stations (BS), radio units (RUs) and devices enable massive levels of cooperation, and increase spatial diversity, which can be exploited for enhancing the radio-sensing performance. The topology is that of a cloud radio access network (C-RAN), where the radio signals are aggregated at the central unit (CU) for joint processing. Although we consider fully centralized processing, this does not preclude some level of distributed processing [14].

The network coverage is divided into serving areas (SA), which are defined as a set of network elements that cooperatively use the same radio resources. Figure 1 illustrates this considering two serving areas. In this case, the resource that distinguishes the SA is the frequency. There is also a certain degree of cooperation between SAs for allocating resources dynamically.

As shown in Figure 1, the infrastructure network elements are linked by the backhaul, for cooperation between them. The infrastructure elements considered include the following: BSs, which are access points with large coverage; RUs, which are access points with limited power and consequently reduced coverage as well as possibly limited processing; and a CU, which represents the location, where the joint processing is performed. The scenario also considers user equipment (UE), namely, transceivers in vehicles and handsets.

Again, the sensing component relies on the concept of cooperative multistatic radar. In this scenario, BSs and RUs within SA1 transmit at carrier F1, while within SA2 they receive at carrier F1; therefore, the two SAs form a multistatic radar. Conversely, if SA2's network elements transmit on carrier F2, and the ones in SA1 receive at that frequency, this forms a second multistatic radar. Furthermore, we assume that the same frequency is used for both sensing and communication. That is, the UEs transmit on the carrier F2 and F1 when they are within SA1 and SA2, respectively. This means that the UEs could also be part of the sensing network as transmitters. If intended to act as receivers, the benefits of cooperation would only be enabled through clusters of cooperating devices, which would require complex exchange of information over the air. However, even as transmitters, the stringent synchronization requirements required for supporting high resolution radio sensing applications would be quite difficult to implement in low-cost devices. Hence we assume that only the infrastructure network elements are involved in the processing. The access of users to radio sensing would be through the client paradigm where the UE requests a service from the infrastructure carrying out the processing and then sends the results.

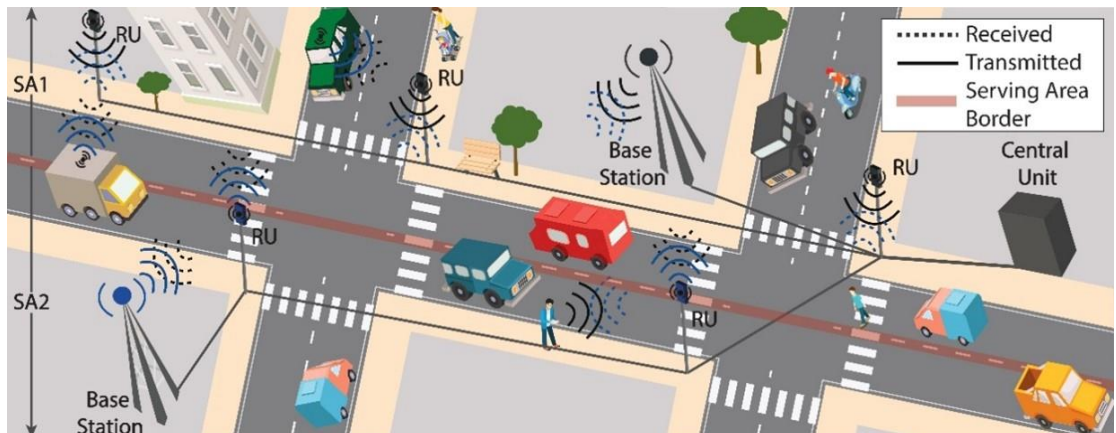

**Figure 1** *RADCOM system architecture for two serving areas controlled by a CU.*

The resource allocation procedures of the transmit / receive bands were not discussed previously, but the trend toward agile flexibility and reconfigurability will continue and this allocation will be dynamic. In the integrated network we will have communication and sensing services, and the resource allocation responsible for the assignment of Tx-Rx nodes and radio resources will be configured dynamically based on the specific demand at a given time / geographical area of both types of services. The frequencies F1 and F2 are not statically defined through an *a priori* planning, but allocated according to the needs. In fact, this could even be supported by a TDD architecture, where a single band exists and the transmission slots of the Tx-Rx units that define the serving area are complementary and the Tx-Rx operations could be reversed.

*Functional Description*

Figure 2 presents the functional diagram of our system proposal, which includes access points, the backhaul and the CU. In the figure we identify the main processing operations required to build the system, while the challenges faced are discussed in the following section. Without any loss of generality, for simplicity, we assume that the transmitters are in SA1 and the receivers are in SA2, but as stated, transmission and reception may occur both ways.

The multistatic radar configuration has $P$ transmitters and $Q$ receivers, connected with the CU through the backhaul for cooperative processing. In Figure 2, there are $PQ$ links, directly defining a distributed MIMO system. Since each access point may be equipped with multiple antennas, we may have a MIMO channel between a transmitter-receiver pair, each having co-located antennas. Each of the $PQ$ links may be viewed as an active bistatic radar component within the multistatic radar configuration, capturing a different bistatic angle and target radar cross-section. The received signals represent the superposition of the signals impinging from the $P$ transmitters, which are forwarded to the CU for joint processing.

From a general perspective, the receiver functions may be divided into three main blocks: synchronization, data processing and channel estimation. In Figure 2, these functions are located at the left of the backhaul, but their location is only indicative, it may vary. Depending on the specific scenario under consideration different types of functionality splitting may be sought. For example, in 5G different functional splitting topologies are under consideration to balance the information flow and processing requirements between the different entities involved, as detailed in [14]. A similar approach may be followed for RADCOM systems.

Regarding the channel estimation and data processing methods, their specific choice is dependent, among other aspects, on the design metric, on the available computational resources, and on the type of information available (local, global, partial, quantized, etc). The methods that are applicable for the communication and sensing functionalities are quite similar. To name just a few, for channel estimation the least squares, maximum likelihood and multiple signal classification techniques are applicable. For data processing, several classic methods may be considered, such as the zero-forcing and minimum mean square error solutions.

The synchronization block operates as a slave of a reference located in the CU. Again, for attaining a high spatial resolution, high-accuracy synchronization is required. Solutions combining Global Navigation Satellite Systems (GNSSs) having high-precision clocks at a few CUs with sophisticated precision time protocols (PTP) are worth investigating. Processing the communication component may remain limited to the physical layer (PHY) or may also include the higher open systems interconnection (OSI) layers [14]. The channel estimation block estimates the CIR of each MIMO link.

However, the radar function typically requires higher delay (range) resolution to cope with the specific requirements of demanding applications dealing with cm/sub-cm object reconstruction. In the same way, tracking an object's movement at a high accuracy requires higher Doppler (velocity) resolution than the one required for communication. This can be achieved by forwarding the received signal to the CU for joint processing.

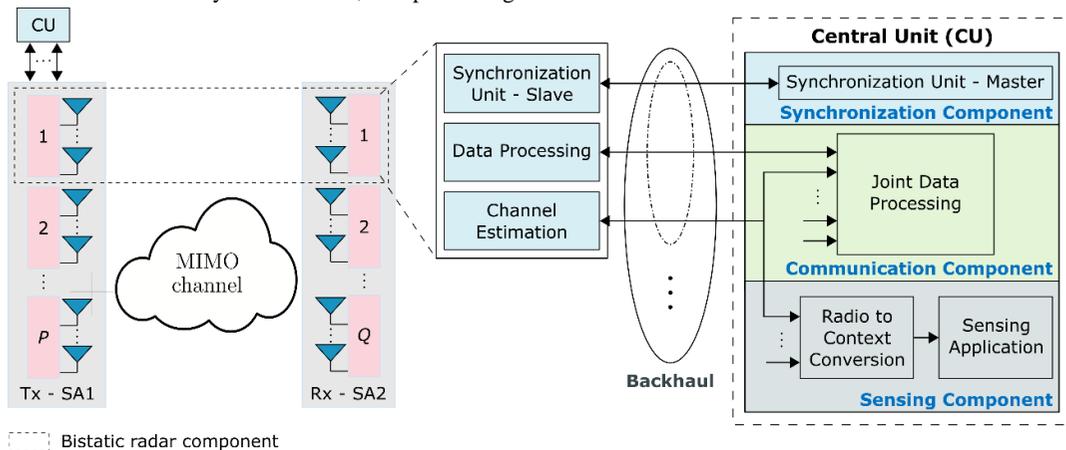

**Figure 2** *Functional diagram of the RADCOM network.*

The CU receives and transmits information from and to all the access points within the RADCOM network's coverage area. In a fully centralized implementation, it aggregates the radio signals, facilitating their joint PHY-layer processing in addition to controlling the operations of the access points. For the proposed RADCOM architecture, we partition the CU operation into three main functional blocks, namely synchronization, communication and sensing components.

The synchronization block is the entity facilitating high-precision-timing-and phase-alignment for attaining high spatial resolution for the sensing component. Again, the CU acts as the master of the synchronization network architecture. The communication component jointly processes the user data from and to the access points within the SAs controlled by the CU.

The sensing component supports situational awareness with the aid of image reconstruction and the tracking as well as identification of objects. The inputs of the CU's sensing component are the $Q$ channel estimates, which will be combined for constructing the CIRs of all $PQ$ channels with an appropriate resolution. According to the proposed algorithm, the accuracy may be improved by exploiting the coded/decoded data from both the communication and radar signals.

The sensing component includes a so-called radio-to-context conversion block. The radar functionality estimates the position, velocity and dimensions of objects/targets. Therefore, it is necessary to translate all radio-related variables into so-called context parameters for the sensing applications. This translation is performed by the radio-to-context conversion block of Figure 2. The output of this block will feed the applications (they may be external to the CU), which will then exploit this context data for reconstructing the environment and tracking objects, for instance.

*Technological Trends*

RADCOM networks would support a wide range of applications and diverse system requirements. For instance, providing wider radio coverage would require the use of sub-6-GHz frequency bands, while high throughputs and high sensing resolutions would call for top-of-the-range technologies, such as massive MIMO (mMIMO) and millimeter wave (mmWave) technologies. Both topics are widely researched by the wireless communications community and are frequently developed together, since the short wavelengths in these bands allow us to accommodate a high number of antennas in terminals. Hence mMIMO-aided mmWave facilitates Gbps throughputs for several concurrent users [6].

Research on these topics in radar is also ongoing [6]. The use of a large number of antennas and the wide bandwidth available at mmWave frequencies allow for high spatial resolution (e.g., 3D reconstruction of targets), enhances the probability of detection, and enables the coexistence of several radar systems [3]. It is anticipated that mMIMO-aided mmWave solutions will be key enablers for RADCOM systems, especially when the sensing application requires high spatial resolution.

The millimeter wave band, which is already under consideration for 5G wireless communication networks. New radio (NR) considers supporting potential use of frequency range up to 100 GHz, enabling bandwidths close to 1GHz (cm level range resolution).

Another key enabling technology to boost the potential of the proposed architecture is machine learning (ML). This technology is already being applied in communications, where the diversity of scenarios, applications, and types of traffic call for adaptive algorithms that do not rely on *a priori* models but learn from data. The intelligent embedding of the additional sensing component will reinforce this need. ML algorithms are data-driven (model-free) instead of being model-based, an important feature for enhancing the system's adaptability and flexibility. For the proposed RADCOM architecture, ML will enhance the cooperative estimation and detection of the algorithms. Additionally, the inclusion of the learning component would improve the speed of the system in terms of parameter acquisition, identification and tracking. The learning component would catalogue and store the relevant context based on previous data that could be fetched in the future.

The need for higher resolution in future applications will evolve towards other technologies, such as the extension of the available spectrum to the THz band. This is already under consideration for 6G systems. As far as our proposal is concerned, the ultra-high bandwidth available for THz carriers would reduce the need for cooperation. However, the accurate channel models are still unknown. Naturally, the path loss will be high, making it only suitable for short range sensing applications, but not for replacing the lower mmWave band.

*Anticipated Applications*

The efficient combination of sensing and communication functions within a common next-generation network enables access to radar-based functionalities. These include surveillance, tracking, ground mapping, environment reconstruction and moving target detection, which may be exploited in multiple areas:

- Intelligent Transportation Systems (ITSs), e.g.
  - the detection of potentially dangerous objects, and
  - the distribution of sensing information among vehicles.
- Internet of Things (IoT), e.g.
  - the sensing of the environment based on reflectometry, and
  - the control of production lines and failure detection in factories.
- Smart cities, e.g.
  - the detection and identification of drones in cities, and
  - the control and management of city facilities such as traffic monitoring, security and vigilance.

Some of the aforementioned applications would not necessarily lead to a generalized deployment using existing systems, mainly due to the high cost of the infrastructure and privacy issues [3]. Regarding the latter point, environmental monitoring using cameras is often viewed as an intrusive method that violates privacy. Furthermore, radar is resilient to low-visibility scenarios, such as dense fog, heavy rain or total darkness. This feature may be crucial for rescue operations in which the transmission of a massive amount of data is also necessary.

## Main Challenges and Foreseen Research Topics

Implementing the proposed RADCOM network faces significant challenges regarding topics such as synchronization, hardware complexity, clutter suppression, and waveform design.

### Network Synchronization

Enhancing radar's resolution implies coherent central processing of the received signals. To achieve this, the local oscillators of all the elements of the network must be time and phase synchronized. In fact, as electromagnetic waves travel 1cm in 33ps, centimeter level resolution accuracy can only be achieved, if the maximum difference between RUs clocks are on the same order of magnitude – again, very tight synchronization is required. As an example, Figure 3 shows the ambiguity function (AF) obtained for a distributed MIMO radar system. The transmitter/receiver antennas are uniformly distributed on the x/y axes with a separation of 2 meters. Figure 3 shows results for two cases, in (a) the signals of all RUs are time and phase synchronized and in (b) they are only time synchronized.

Figure 3 shows that the resolution is two orders of magnitude better in the perfectly synchronized case (observe that the x and y scales in Figure 3 (a) and (b) are different).

There are several synchronization protocols and algorithms, such as PTP, master-slave closed-up, round trip, and broadcast consensus algorithms [15]. Although these can meet the more relaxed synchronization requirements of the operational MIMO radar systems, the future RADCOM requirements are beyond what the existing solutions can provide. Dedicated synchronization solutions based on the master slave-concept and extending the concept of the phase-lock loop to distributed environments will be needed.

### Hardware Complexity

Again, mMIMO-aided mmWave techniques allow for packing a large number of antennas in a small form-factor, facilitating portability and allowing the exploitation of spatial processing techniques, such as beamforming [3]. However, the hardware limitations of high-frequency carriers lead to several challenges regarding both the complexity and implementation costs [8]. Hybrid analog-digital beamforming (HAD) and sub-arrayed MIMO radar mitigate both the complexity and cost of RADCOM [6].

### Clutter Suppression

Multistatic radar relies on reflections for detecting, identifying and tracking objects. However, the large number of noise-like backscattered reflections encountered in UDN scenarios will act as increased clutter in radar jargon, hence degrading the radar performance. The clutter in UDN environments would be mainly comprised of the superposition of reflections. For conventional radar systems, clutter suppression is a widely studied topic, which mitigates backscattering resulting from the surface of sea and ground. It is clear that the scenario investigated in this study is quite different from the traditional ones, which requires radical new research on clutter suppression [11]. Machine learning techniques constitute an important tool of addressing this problem, since we cannot predict all environmental scenarios and therefore are unable to build models for all possible situations.

### Waveform Design

The design of the waveform is crucial for the development of RADCOM systems. For instance, MIMO radar employs a set of orthogonal and deterministic signals having good correlation properties. On the other hand, MIMO communication systems may rely on quasi-orthogonal signals and random modulated user data. Furthermore, both single carrier and multicarrier communication waveforms have to obey design trade-offs in RADCOM systems [2]. OFDM has been widely investigated as a promising integrated paradigm because of its convenient frame structure, simplicity, and parametrization possibilities. However, its high PAPR is inherited under the RADCOM paradigm. Therefore, the single carrier cyclic prefix (SCCP) aided waveform has also been considered for RADCOM, since it exhibits lower PAPR than OFDM at a similar communication performance, but at a potentially higher complexity. Another design option is to employ the orthogonal time-frequency space (OTFS) waveform, which has been shown to be more suitable than OFDM for high-mobility scenarios. In spite of waveform designs have recently been conceived for RADCOM systems, further research is required, especially when considering joint mmWave and mMIMO systems.

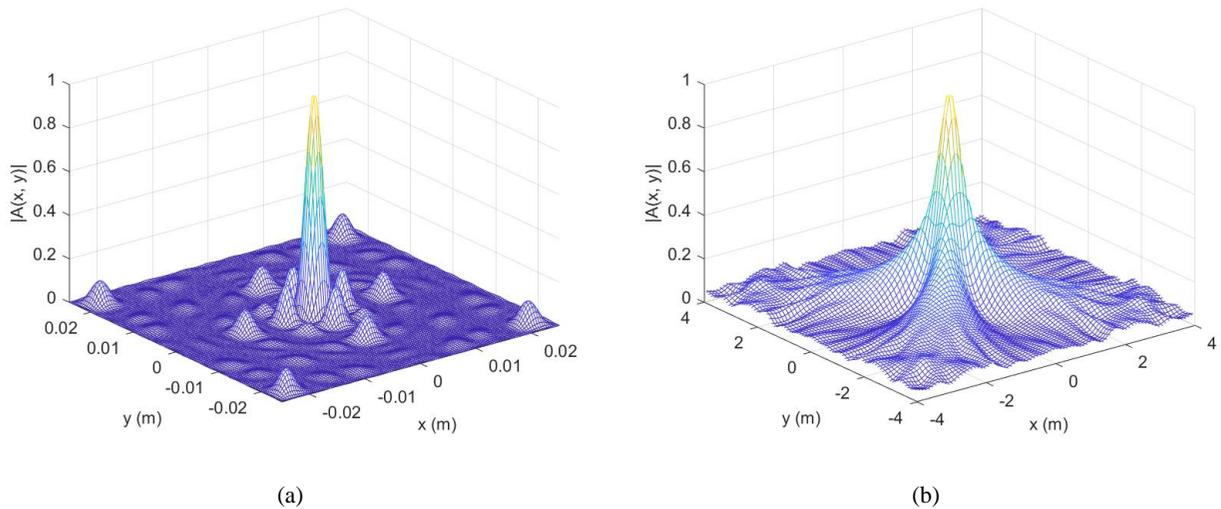

**Figure 3** *Two-dimensional ambiguity function: (a) synchronized and (b) unsynchronized. Parameters: carrier frequency 26 GHz; OFDM with 32 subcarriers; bandwidth 400 MHz; SNR 30 dB.*

### *RADCOM's optimal Pareto-front*

Similarly to the widely known 'spider-diagram' of the 5G systems, which portrays the set of conflicting design trade-offs of throughput, latency, bit error rate (BER), power-efficiency etc, future RADCOM systems have to rely on multi-component optimization. More explicitly, instead of for example maximizing the throughput of a RADCOM system at the cost of degrading its power-efficiency or BER or localization accuracy, a community-effort is required for finding the entire Pareto-front of all optimal solutions. Once all the optimal solutions of the Pareto-front have been found, the most appropriate configuration may be activated, as exemplified by the ultra-reliable low latency communications (URLLC) or enhanced mobile broadband (eMBB) 5G modes, which have very different characteristics. However, none of the Pareto-optimal operating points allow the improvement of say the BER without degrading the power-efficiency or without extending the latency for example.

## Conclusions

In this article, we proposed a framework that includes a sensing component in UDN networks. This integration will enable access to new radar-based applications for users. We introduced the overall system architecture and described the components of the RADCOM network, namely, the multistatic radar configuration as well as its receiver, and the CU. Several technological trends required for achieving high spatial resolution, improved detection and precise tracking capabilities were identified, such as mMIMO-aided mmWave and ML solutions. The road leading to the joint sensing-communication network paradigm has significant challenges. Namely, the network synchronization, clutter suppression, hardware complexity, and waveform design still require significant research. The article also identified compelling applications. The proposal is limited to fixed elements of cellular networks, but the paradigm may be extended to mobile scenarios, where moving network elements are also part of the multistatic radar. This extended paradigm would include scenarios such as autonomous driving and networks of unmanned aerial vehicle.

## Acknowledgments

This work has received funding from the European Union's Horizon 2020 research and innovation programme under the Marie Skłodowska-Curie ETN TeamUp5G, grant agreement No. 813391, and from the FCT/MCTES through national funds and when applicable co-funded EU funds under the project UIDB/50008/2020-UIDP/50008/2020. L. Hanzo would like to acknowledge the financial support of the Engineering and Physical Sciences Research Council projects EP/P034284/1 and EP/P003990/1 (COALESCE) as well as of the European Research Council's Advanced Fellow Grant QuantCom (Grant No. 789028)

## Author Information

*Leonardo L. Lamas* (leoleval@av.it.pt) received an M.Sc. degree in Electronic Engineering with an emphasis on Communication Systems from the Pontifical Catholic University of Rio de Janeiro in 2018. He is currently pursuing a Ph.D. degree at the University of Aveiro. Since 2019, he is an early-stage researcher at the TemUp5G ETN, conducting his actual research at the facilities of Instituto de Telecomunicações – Polo Aveiro (IT-AV) Aveiro. His research interests include MIMO radar, wireless communication systems, and Integrated Sensing and Communication (ISAC) paradigm.

*D. Castanheira* (dcastanheira@av.it.pt) received Licenciatura (ISCED level 5) and Ph.D. degrees in electronics and telecommunications from the University of Aveiro in 2007 and 2012, respectively. He is currently an auxiliary researcher at the Instituto the Telecomunicações, Aveiro, Portugal. He has been involved in several national and European Projects, namely, RETIOT, SWING2, PURE-5GNET, HETCOP, COPWIN, and PHOTON within the FCT Portuguese National Scientific Foundation and CODIV, FUTON and QOSMOS with the FP7 ICT. His research interests are signal processing techniques for digital communications with an emphasis on physical layer issues including channel coding, precoding/equalization and interference cancelation.

*Adão Silva* (asilva@av.it.pt) received M.Sc. and Ph.D. degrees in Electronics and Telecommunications from the University of Aveiro in 2002 and 2007, respectively. He is currently an assistant professor in the DETI at the University of Aveiro and a senior researcher at the Instituto de Telecomunicações. He has participated in several national and European projects. He has led several research projects in broadband wireless communications at the national level. He served as a member of the TPC at several international conferences. Currently, he is an associate editor of IEEE Access and IET Signal Processing. His interests include multicarrier-based systems, cooperative networks, precoding, multiuser detection and massive MIMO.

*Atílio Gameiro* (amg@ua.pt) received Licenciatura and Ph.D. from the University of Aveiro in 1985 and 1993, respectively. He is currently an associate professor in the Department of Electronics and Telecom. at the University of Aveiro and a researcher at the Instituto de Telecomunicações, where he is the Head of the group. His main interests are in signal processing techniques for digital communications and communication protocols. He has published over 200 technical papers in international journals and conference proceedings. His current research activities involve space-time-frequency algorithms for broadband wireless systems and cross-layer design. He has been involved and has led the participation of the IT or University of Aveiro in more than 20 national and European projects.

*Lajos Hanzo* (http://www-mobile.ecs.soton.ac.uk, https://en.wikipedia.org/wiki/Lajos/_Hanzo) (FIEEE'04) received his Master degree and Doctorate in 1976 and 1983, respectively from the Technical University (TU) of Budapest. He was also awarded the Doctor of Sciences (DSc) degree by the University of Southampton (2004) and Honorary Doctorates by the TU of Budapest (2009) and by the University of Edinburgh (2015). He is a Foreign Member of the Hungarian Academy of Sciences and a former Editor-in-Chief of the IEEE Press. He has served several terms as Governor of both IEEE ComSoc and of VTS. He has published 1970 contributions at IEEE Xplore, 19 Wiley-IEEE Press books and has helped the fast-track career of 123 PhD students. Over 40 of them are Professors at various stages of their careers in academia and many of them are leading scientists in the wireless industry. He is also a Fellow of the Royal Academy of Engineering (FREng), of the IET and of EURASIP.